\documentstyle[12pt,psfig]{article}

\def\lesssim{\mathrel{\mathpalette\vereq<}}
\def\gtrsim{\mathrel{\mathpalette\vereq>}}
\makeatletter
\def\vereq#1#2{\lower3pt\vbox{\baselineskip1.5pt \lineskip1.5pt
\ialign{$\m@th#1\hfill##\hfil$\crcr#2\crcr\sim\crcr}}}
\makeatother

\def\c#1{\setbox0=\hbox{#1}\ifdim\ht0=1ex \accent'30 #1%
	\else{\ooalign{\hidewidth\char'30\hidewidth\crcr\unhbox0}}\fi}

\begin{document}
\begin{titlepage}
\begin{center}
~ \hfill    LBL-39030\\
~ \hfill UCB-PTH-96/28\\

\vskip .25in
\renewcommand{\thefootnote}{\fnsymbol{footnote}}
{\large \bf Excluding Light Gluinos from $Z$ decays}\footnote{AdG was
supported by CNPq (Brazil).  HM was supported in part by the Director,
Office of Energy Research, Office of High Energy and Nuclear Physics,
Division of High Energy Physics of the U.S. Department of Energy under
Contract DE-AC03-76SF00098 and in part by the National Science
Foundation under grant PHY-95-14797.}

\vskip 0.3in

Andr\'e de Gouv\^ea and Hitoshi Murayama\\
~\\
{\em Department of Physics,
     University of California\\
     Berkeley, California 94720}\\
and\\
{\em Theoretical Physics Group\\
     Ernest Orlando Lawrence Berkeley National Laboratory\\
     University of California,
     Berkeley, California 94720}
\end{center}

\vskip .3in

\vfill

\begin{abstract}
We reanalyze the constraints on light gluinos ($m_{\tilde{g}}\leq
5$~GeV$/c^2$) from the 
hadronic $Z$ decays into four jets.  We find that the published OPAL data 
from the 1991 and 1992 runs exclude a light quasi-stable gluino with
mass $\lesssim 
1.5$~GeV$/c^2$ at more than 90\% confidence level.  This limit depends
little on assumptions about the gluino fragmentation and the definition of
the gluino mass.  The exclusion confidence 
level is shown as a function of the mass.  A future projection is 
briefly discussed.  We also discuss quantitatively how the 
distributions in the Bengtsson--Zerwas and the modified
Nachtmann--Reiter angles change due to the finite bottom quark or gluino
mass.
The analysis is limited to the leading-order calculations.
We, however, give an empirical reason to why the next-to-leading-order
corrections are unlikely to change our conclusions.
\end{abstract}

\vfill

\end{titlepage}

\renewcommand{\thepage}{\arabic{page}}
\setcounter{page}{1}

Supersymmetry is one of the primary targets of extensive searches at 
various collider experiments, most importantly at CERN $e^{+}e^{-}$ 
collider LEP and Fermilab $p\bar{p}$ collider Tevatron \cite{DPF}.  
Negative searches at these and previous colliders have already put 
significant constraints on the parameter space of low-energy 
supersymmetry.  However, a light gluino below the few GeV mass range 
has surprisingly weak experimental constraints as emphasized recently by 
various authors \cite{Clavelli1,FarrarPRD,Barnett} (see, however, an 
opposing view \cite{Haber}).  It is an extremely important task to 
verify or exclude a gluino in this light window experimentally.  While 
the Tevatron Run II is expected to extend the reach of heavy gluinos 
up to a few hundred GeV, little effort is devoted to 
definitively exclude or verify the light gluino window.  On the other 
hand, a careful reexamination of the existent data may reveal an 
overlooked constraint on a light gluino; this is our 
motivation to study the existent data in detail.

We reanalyzed published data on $Z$ decays into four jets 
\cite{L3,OPAL1,DELPHI1,ALEPH,DELPHI2,OPAL2}, and found that
they already exclude a gluino lighter than 1.5~GeV$/c^2$ at more than
90\% confidence level.  We assume that the gluino does not decay inside
the detector. 
Since the published results use only 1991 and 1992 data, it is
conceivable that the currently available data, if analyzed properly,
could put a much more significant constraint on a light gluino.  We hope
our result urges the experimental groups to
analyze the whole data set including a possible light gluino.

Let us briefly review the existent constraints on a light gluino (see
\cite{FarrarPRD,PDG} for more details).  The negative searches at beam dump
experiments have excluded a light gluino which decays inside the
detector into photino, which in turn interacts with the neutrino
detector.  However, a gluino tends to leave the
detector without decaying if the squark mass is above a few hundred
GeV$/c^2$ \cite{FarrarPRL,longlived}.  Even if the gluino decays, the photino
interacts very 
weakly in this case and cannot be detected.  If the gluino does not
decay, it forms bound states such as gluinoball $\tilde{g}\tilde{g}$,
glueballino $g\tilde{g}$ or baryon-like states, especially $uds\tilde{g}$
\cite{udsgluino}.  Other states are likely to decay into these
neutral 
bound states, and searches for exotic charged hadrons may not apply
unless a charged gluino bound state decays only weakly.  
One the other 
hand, the mass region above 1.5~GeV$/c^2$ and below 4~GeV$/c^2$ is excluded 
from quarkonium decay $\Upsilon 
\rightarrow \gamma \eta_{\tilde{g}}$, where $\eta_{\tilde{g}}$ 
is the pseudo-scalar gluinoball,
independent of the gluino lifetime \cite{CF,FarrarPRD}.  Whether the bound
extends to lower masses is  
controversial because of the applicability of perturbative QCD 
calculations \cite{CF}.  The mass range above 4~GeV$/c^2$ is expected to
give a  
shorter lifetime and is excluded by a negative search for events with 
missing energy at UA1 \cite{UA1}.  The authors of \cite{BHK} claim 
that the limit from UA1 extends down to 3~GeV$/c^2$.  In any case, the 
least constrained region is the mass range below 1.5~GeV$/c^2$, where the 
gluino is relatively stable so that it does not decay inside 
detectors.  This is our window of interest in this letter.  

We would like to emphasize that the best method to exclude the gluino 
mass range below 1.5~GeV$/c^2$ is to use inclusive processes rather than 
searching for specific bound states with certain decay modes.  The 
latter search would heavily depend on assumptions such as the mass 
spectrum of various gluino bound states and their decay modes and 
decay lifetimes.  One would have to design experiments and put constraints 
with all possible theoretical assumptions on gluino bound states in 
order to exclude the light gluino definitively.  On the other hand, the 
constraints would be much less sensitive to theoretical assumptions if 
they were based on inclusive processes where perturbative QCD is 
applicable.  There are several possibilities pointed out in the literature 
along this line.  The most popular one is to study the effect of light 
gluinos in the running of the QCD coupling constant $\alpha_{s}$.  It was 
even pointed out that the values of $\alpha_{s}$ from higher energy 
measurements tend to be higher than those extrapolated from lower 
energies using QCD with the ordinary quark flavors, and the data actually 
prefer the existence of a light gluino to compensate the slight 
discrepancy \cite{Clavelli1,Clavelli2,Clavelli3}.  However, 
this issue remains controversial \cite{Antoniadis,Hebbeker,ENR}.  
Even though the discrepancy between low-energy and high-energy 
measurements is diminishing \cite{HinchliffePDG}, still the data are 
not precise enough to exclude or verify a light gluino definitively.  
The second one is its effect on the Altarelli--Parisi evolution of the 
nucleon structure functions \cite{RS,Blumlein}.  Unfortunately the effect 
is too small to be tested using the present experimental data.  It might be 
that the more recent HERA data could improve the situation, but making a
definite statement on the existence of a light gluino appears to be
difficult.   The third one is to study the angular correlations in the
so-called ``3+1'' jet events at HERA \cite{MunozHERA}.  However, the
effect of the light gluino was found to be negligible.
The final one, which we employ in this letter, is the study of four jet 
correlations in $e^{+} e^{-}$ collisions 
\cite{Farrar4jet,ENR,Munoz4jet}.  Previous studies did not find
significant constraints, but given the size of the current LEP data,
we find this to be the most promising direction.

The only data we use in this letter are studies of QCD color factors 
\cite{ALEPH,DELPHI2,OPAL2}.  The experimental groups at LEP have 
performed impressive analyses of the hadronic $Z$ decays into four jets, 
extracting QCD color factors $C_{A}/C_{F}$ and $T_{F}/C_{F}$
\cite{color} from jet 
angular distributions, to confirm SU(3) as the QCD gauge group and five light 
quark flavors.  The angular distributions of $q\bar{q}q\bar{q}$ final 
state differ from those of $q\bar{q}gg$, where $q$ refers to a generic 
quark and $g$ to a gluon.  Three angles are commonly used in four-jet
analyses: the Bengtsson--Zerwas (BZ) angle $\chi_{BZ}$ \cite{BZ}, the
modified Nachtmann--Reiter (NR) angle $\theta^*_{NR}$ \cite{NR}, and the
opening angle of the two less energetic jets $\alpha_{34}$.  
If there exists a light gluino $\tilde{g}$, 
the final state $q\bar{q}\tilde{g}\tilde{g}$ also contributes to the 
$Z$ decays into four jets.  The angular distributions of 
$q\bar{q}\tilde{g}\tilde{g}$ would be identical to those of 
$q\bar{q}q\bar{q}$.  Therefore, a possible light gluino would change 
the extracted $T_{F}/C_{F}$ but not $C_{A}/C_{F}$.  Apart from the mass
effects, $T_{F}/C_{F}$ should increase by a factor of $(5+3)/5$, because the 
gluino is a color-octet and counts effectively as three additional 
massless quarks.  Note that these analyses do not use the overall
rate of four-jet events since it is sensitive to the choice of
$\alpha_s$ in the absence of next-to-leading order (NLO) calculations.
So far the experimental analysis which used the 
highest statistics is the one by 
OPAL \cite{OPAL2}, which also briefly discussed 
constraints on a light gluino.  They found that the light gluino is 
barely outside the 68\%
confidence level contour and decided the data did not put a significant 
constraint.

However, we find the previous analyses not carefully designed to study 
the effect of a light gluino because of the following reason.  When 
one discusses a possible light gluino, QCD with the color group SU(3) 
should be assumed.  Given overwhelming experimental evidences of QCD, 
it is not wise to, for instance, vary the number of colors $N_{c}=3$ 
when one studies the effect of a particle (light gluino) {\it added 
to}\/ QCD. Therefore, we must fix the QCD color factor $C_{A}/C_{F}$ 
to be that of the SU(3) group, 9/4.  Second, we already {\it know}\/ that 
there are five quark flavors $u$, $d$, $s$, $c$ and $b$, which appear 
in $Z$ hadronic decays.  When one puts constraints on an {\it 
additional}\/ contribution from a light gluino, one should not vary 
the number of flavors below 5, or equivalently, $T_{F}/C_{F}$ below 3/8.
The only LEP paper which analyzed data 
in a way close to this spirit, and put an upper bound on possible 
additional $q\bar{q}q\bar{q}$-type final states, is the one from OPAL 
\cite{OPAL1}; but it used very limited statistics.  All more recent 
papers \cite{ALEPH,DELPHI2,OPAL2} varied both $C_{A}/C_{F}$ and 
$T_{F}/C_{F}$ without constraints.  By reanalyzing data with these 
constraints we can put a much more significant bound on a light gluino 
than reported.  Actually, fixing the group to be SU(3)
($C_{A}/C_{F}=9/4$) has the greatest impact on the 
confidence level, while restricting $T_{F}/C_{F}\geq 3/8$ has a much smaller
effect 
(actually it makes the significance worse).  We further include the 
finite mass of the bottom quark in the analysis which slightly improves the 
significance.  Overall, a massless gluino is excluded already better 
than at 90\% confidence level by the OPAL 1991 and 1992 data only
\cite{OPAL2}. 

\begin{figure}[t]
\centerline{
\psfig{file=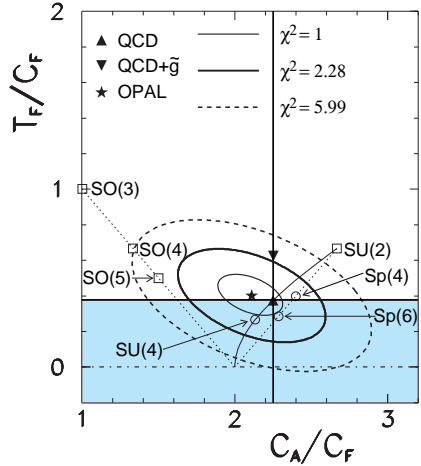,width=0.4\textwidth}
}
\caption[1]{Extracted QCD color factors from the OPAL analysis
	\cite{OPAL2}.   The shown $\chi^2$ values correspond to 39.3\%,
	68\% and 95\% confidence levels with two degrees of freedom.
	We impose the constraint $C_{A}/C_{F} = 9/4$ (vertical solid
	line) and limit ourselves to the unshaded region ($T_{F}/C_{F} \geq 
	3/8$) in order to put constraints on a possible light gluino 
	contribution to the four-jet events from $Z$ decays.
	See the text for more details.
}
\end{figure}

Let us start from the reported contour on the $C_{A}/C_{F}$, $T_{F}/C_{F}$ 
plane, shown in Fig.~1.  We fix $C_{A}/C_{F} = 9/4$ because of the 
philosophy of our study stated above.  Since one-dimensional 
$\chi^{2}$ distributions have much higher confidence levels than 
two-dimensional ones, this change improves the significance of the data 
drastically.  From their $\chi^{2}$ contours, we minimized $\chi^{2}$
with fixed $C_{A}/C_{F}=9/4$, and defined $\Delta \chi^{2}$ relative to
the $\chi^{2}$ at the minimum ($T_{F}/C_{F} = 0.36$).  The confidence
levels are calculated using a one-dimensional $\chi^{2}$ distribution with
$\Delta \chi^{2}$ defined in this manner.  This is a conservative choice
because $\Delta \chi^{2} < \chi^{2}$.  
We obtain $T_{F}/C_{F} = 0.36 \pm 0.15$ with fixed 
$C_{A}/C_{F}$.  If one had used this central value and 
the standard deviation, a massless gluino would be excluded at 95\% confidence 
level.  However, we also need to impose another constraint, $T_{F}/C_{F}
\geq 3/8$, which can be easily taken 
into account.  The standard method is to use the Gaussian distribution 
only in the physical region, and scale the normalization of the 
distribution so that the total probability in the physical region 
becomes unity.  Since the central value is very close to the 
theoretical value of the QCD, this effectively increases the probability 
of allowing light gluinos by a factor of two; numerically the 
confidence level is 88\%.

Finally, we study the effect of the finite mass of the bottom quark and 
gluinos on the extracted $T_{F}/C_{F}$.  The authors of \cite{BMM} 
studied the effect of the finite mass of quarks on the four-jet rates.  
They also looked at the angular distributions and reported there were 
little changes.  Even though it is true that the distributions do not 
change drastically, they gradually become similar to those of 
$q\bar{q}gg$ final state as one increases the mass of the quark, and 
hence the extracted $T_{F}/C_{F}$ from the fit to the distributions 
has a relatively large effect due to the finite mass of the bottom 
quark.  
The papers \cite{ALEPH,DELPHI2} do not take this effect into account at 
all.  The OPAL experiment \cite{OPAL2} used parton level event 
generators by the authors of \cite{BMM} and \cite{Munoz4jet} to study the 
effect.  They have found a surprisingly large effect: the bottom quark 
contribution to $T_{F}/C_{F}$ was about one half of a massless quark at 
$y_{cut}=0.03$.  We confirmed their estimate in a detailed 
parton-level calculation based on that done in \cite{baryon}, neglecting the 
interference between primary and secondary quarks.  This  
approximation is known to be better than a few percent.  On the other 
hand, this approximation has the clear advantage of enabling us to 
distinguish primary and secondary quarks unambiguously.  Our code 
employs helicity amplitude technique using the HELAS package \cite{HELAS}, 
which made it straight-forward to incorporate finite masses in the 
four-jet distributions.

\begin{figure}[t]
\centerline{
\psfig{file=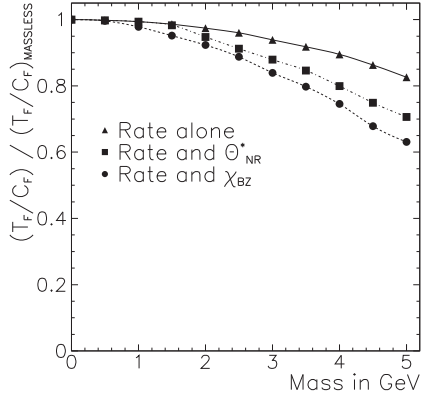,width=0.4\textwidth}
}
\caption[2]{Effective contribution to $T_{F}/C_{F}$ of a massive 
	secondary quark relative to the massless case.  The solid line 
	shows the reduction in the rate alone.  The other two lines 
	include the effect that the distributions in BZ and NR angles 
	change due to finite quark mass.  We chose $y_{cut}=0.03$ and
	$\sqrt{s} = m_Z c^2 = 91.17$~GeV.}
\end{figure}

The finite mass affects the extracted $T_{F}/C_{F}$ in two ways.  
First, the rate of producing secondary massive quarks is suppressed 
compared to the massless case as shown with the solid line in 
Fig.~2.   For instance, there is 
about 20\% suppression with $m_{q} = 5$~GeV$/c^2$ and $y_{cut} = 0.03$.
This result is consistent with \cite{BMM}.  The mass of the 
primary quark has little effect on the rate: only a 6\% suppression for $m_{q} 
= 5$~GeV$/c^2$. We also checked that the distributions in BZ and NR
angles with a massive primary quark
are indistinguishable from the massless case.  These observations are 
consistent with naive expectations, because the primary quarks are much 
more energetic than the secondary ones and hence the mass effect is 
suppressed by $m^{2}/E^{2}$.  We therefore neglect the finite mass of primary 
quarks hereafter.  Second, the NR and BZ angle distributions
gradually approach those of the $q\bar{q}gg$ final state as one increases 
the mass of the secondary quarks.  We are not aware of detailed analyses 
of these distributions with massive quarks in the literature.  The 
distributions are shown in Fig.~3 normalized so that the total area below 
the curve is unity, in order for the effect on the rate and that on the 
distribution to be clearly separated.  We fit the distributions as 
linear combinations of $q\bar{q}gg$ and massless $q\bar{q}q\bar{q}$ 
distributions to determine the {\it effective}\/ $T_{F}/C_{F}$, in 
order to mimic the experimental analyses.  The fit is surprisingly good; 
we checked this for quark masses between 0 and 5~GeV$/c^2$.  Combined with the 
reduction in the rate, the net effect of the finite mass of secondary 
quarks is shown in Fig.~2.  With $m_{b}=5$~GeV$/c^2$ for 
secondary bottom quarks, the overall rate of $q\bar{q}b\bar{b}$ final state 
is reduced to 82.5\%, while the fit to angular distributions gives a 
$T_{F}/C_{F}$ reduced to 76.4\% (BZ) or 85.5\% (NR) compared to that of a 
massless quark flavor (3/8), on top of the reduction in the rate.  In total, 
secondary bottom quarks contribute to $T_{F}/C_{F}$ as $3/8*0.630$ or 
$3/8*0.705$, which is not a negligible suppression.  The extracted
$T_{F}/C_{F}$ from the data is an average of $T_{F}/C_{F}$ from five
flavors.  The reported 
$T_{F}/C_{F}$ in \cite{OPAL2} includes a correction to compensate the apparent 
suppression due to the finite bottom quark mass.  Such a correction in 
turn effectively enhances the additional contribution from gluinos by a 
factor of $5/(4+0.630)$ or $5/(4+0.705)$.  Note that this slight
enhancement effect does not change significantly even when one varies $m_b$
from 4 to 5~GeV$/c^2$, as can be seen in Fig.~2.

\begin{figure}[t]
\centerline{
\psfig{file=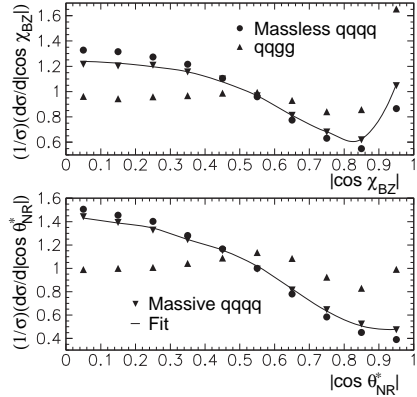,width=0.4\textwidth}
}
\caption[3]{The distributions in BZ and NR angles of the 
	$q\bar{q}q\bar{q}$ final state where the secondary quark has a 
	mass of 5~GeV$/c^2$.  They can be fit extremely well as a linear 
	combination of massless $q\bar{q}q\bar{q}$ and $q\bar{q}gg$ 
	distributions.  We used $y_{cut}=0.03$ and $\sqrt{s}=m_Z c^2$.}
\end{figure}

The actual OPAL analysis \cite{OPAL2} fits the data in the 
three dimensional space spanned by BZ, NR and 
$\alpha_{34}$ angles after bin-by-bin systematic corrections from Monte Carlo 
simulations.  Such an analysis is beyond the scope of this letter.  
We assume that the total effect of the 
finite mass is somewhere between the effects on BZ or NR angles since 
$\alpha_{34}$ is not as effective in extracting $T_{F}/C_{F}$.  
As it is clear from Fig.~3, fits to distributions of massive quarks 
give apparent additional contributions to $q\bar{q}gg$ and hence 
$C_{A}/C_{F}$.  They are completely negligible, however, compared to 
the size of the true $q\bar{q}gg$ which is about one order of magnitude 
larger than the sum of all $q\bar{q}q\bar{q}$ final states, and hence 
we will neglect such contributions hereafter.  

\begin{figure}[t]
\centerline{
\psfig{file=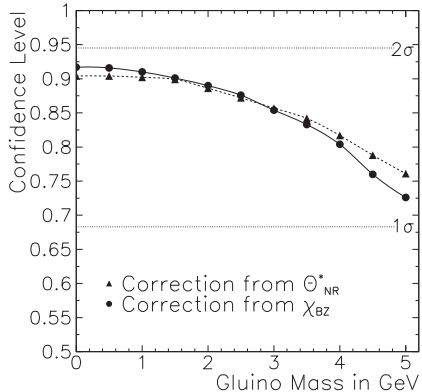,width=0.4\textwidth}
}
\caption[4]{Exclusion confidence level of a light gluino as a function 
	of its mass.  Two curves are shown depending on the method of 
	estimating the finite mass effects.  In either case, a light 
	gluino of mass below 1.5~GeV$/c^2$ is excluded at more than 90\% 
	confidence level.
}
\end{figure}

Given the above considerations, we can now present the exclusion 
confidence levels on a light gluino for varying gluino masses in 
Fig.~4.  For both curves, we used $m_{b} = 5$~GeV$/c^2$ and used the 
effective $T_{F}/C_{F}$ extracted from the fits to BZ and NR angles.  
The finite mass effect of the gluino is treated in the same manner.  
First of all, it is clear that the finite mass effect which we studied 
depends little on the choice of BZ or NR angles, and hence we believe 
it mimics the true experimental fits (which use BZ, NR and 
$\alpha_{34}$ angles simultaneously in a three-dimensional fit with 
295 bins) quite well.  Second, the confidence level is extremely flat 
up to 2~GeV$/c^2$.  This implies that we do not need to worry about 
complication due to non-perturbative dynamics in defining the gluino 
mass \cite{mq}.  The lower bound of $\simeq 1.5$~GeV$/c^2$ at 90\% 
confidence level is already 
in the perturbative region.  It is quite likely that the gluino mass 
relevant to this analysis is a running mass defined at the scale $Q^2 
\sim y_{cut} m_Z^2$ \cite{Ellis}.  It is then straight-forward to convert 
the bound to the on-shell gluino mass: the lower bound of
$\bar{m}_{\tilde{g}} (0.03 m_Z^2) = 1.5$~GeV$/c^2$ in the $\overline{\rm
MS}$ scheme  
corresponds to $m_{\it pole} (\tilde{g}) = 2.8$~GeV$/c^2$. 

We would like to comment that the clever jet reconstruction method used
in the OPAL analysis \cite{OPAL2} is particularly suited for the study
of light gluinos in four-jet events.  They did not scale the measured
jet energies by an overall ratio $E_{vis}/m_{Z}$, as done traditionally
in similar analyses, but instead used the angular information of the
jets to calculate the energy of each jet using energy and momentum
conservation.  This method avoids uncertainties in the gluino fragmentation.
Since it is not well understood how a gluino fragments, one should use a
similar method to avoid dependence on assumptions about the gluino
fragmentation in future studies.


Unfortunately, our analysis is limited to the leading-order (LO)
calculations.  It is a natural question whether the
NLO corrections may change our conclusion.  First of all, we expect that the
corrections to the angular variables used in the analysis are presumably
not large.  The NLO corrections are important when a variable
involves $\alpha_s$, such as 3- and 4-jet rates, thrust, etc.
The variables used in our analysis are not proportional to powers of
$\alpha_s$, and hence scale-independent at the LO
approximations.  This is analogous to the case of the forward-backward
asymmetry which is an (integrated) angular variable and is $\alpha_s$
independent at the LO.  It does receive an NLO correction
of $c (\alpha_s/\pi)$, where $c \sim 0.89$ in
the case of a massless quark \cite{DLZ}.  In our case, we also expect a
correction to the angular distributions of the order of
$\alpha_s(\mu)/\pi$, where $\mu^2 \sim y_{cut} m_Z^2$ is probably an
educated guess.  Then a typical size of the NLO
correction is about 5~\%.  However, a correction of this
order of magnitude may still be of concern because of the following
reason.  The $q\bar{q}gg$ final state is roughly an order of magnitude
larger than the $q\bar{q}q\bar{q}$ final state.  Therefore, a 5\%
correction to $q\bar{q}gg$ may result in a 50\% correction to
$q\bar{q}q\bar{q}$ final state, to be compared with a possible 60~\%
contribution from the gluino.  

We argue, however, that such a higher
order correction is not likely to change our conclusion.  First of all,
the helicity structure and the color flow in the $q\bar{q}gg$ final state
and $q\bar{q}q\bar{q}$ 
final state are quite different.  If a correction to the $q\bar{q}gg$
final state changes the conclusion, the following must be
happening: the correction term to the $q\bar{q}gg$ exactly mimics an
additional contribution to the $q\bar{q}q\bar{q}$ final state in the
angular distributions with a negative sign such as to mask the
contribution from the $q\bar{q}\tilde{g}\tilde{g}$ final state.  We do
not find this to be likely because they have different structures in the
helicities and colors.
Moreover, the data do not indicate that the NLO correction
is large.  OPAL data \cite{OPAL2} are fit very well by the LO
Monte Carlo on three-dimensional histograms of 295 bins with
$\chi^2/{\rm d.o.f} =290/292$.  
This excellent agreement between the matrix element calculation and the data
found in \cite{OPAL2} supports the smallness of the
NLO corrections empirically.  However, the calculations
of NLO corrections are necessary to justify it.\footnote{It is
encouraging that partial NLO calculations were done after the completion
of this work \cite{NLO}.  A preliminary study shows that the
correction from leading terms in $1/N_c^2$ expansion is small \cite{Dixon}.}
For future studies, it is also desirable to 
compare different Monte Carlo programs, while only JETSET was used in
recent experimental papers \cite{ALEPH,DELPHI2,OPAL2}.

Finally, it is worth emphasizing that the result in this letter is 
based on the 1991 and 1992 OPAL data with 1.1M hadronic $Z$'s 
\cite{OPAL2}.  The statistical and systematic uncertainties are 
comparable in their paper.  Given the current size of the LEP data, which 
is more than an order of magnitude larger, the statistical uncertainty 
should reduce substantially once all of the data has been analyzed.  
This change alone could drastically improve the sensitivity to the 
light gluino in four-jet events.  On the other hand, it is not obvious 
how systematic uncertainties can be further reduced.  The largest 
systematic uncertainty originates in the bin-by-bin acceptance 
corrections which needed to be done before performing a fit in BZ, NR, 
and the opening angle space.  It is not clear how this uncertainty can be 
reduced if one employs the same method.  Perhaps choosing larger values 
of $y_{cut}$ reduces the uncertainty while reducing the statistics at 
the same time.  There could be an optimal choice of $y_{cut}$ for this 
particular purpose.  Some of the other large systematic uncertainties are 
specific to the OPAL experiment and could be reduced by averaging results 
from all four experiments.  In any case, there is no doubt that we can 
expect a better result from the currently available data set.

In summary, we reanalyzed the published OPAL 1991 and 1992 data on the
QCD color factors \cite{OPAL2} to constrain possible additional
contributions to four-jet events in $Z$ decays due to
$q\bar{q}\tilde{g}\tilde{g}$ final states.  The main difference from the
original OPAL study is to fix $C_{A}/C_{F} = 9/4$ as required by QCD.
We further imposed $T_{F}/C_{F} \geq 3/8$ and treated the finite mass
effects of both the bottom quark and the gluino carefully.  We find that
a light gluino with a mass below 1.5~GeV$/c^2$ is excluded at better
than 90\% confidence level.  
The result is insensitive to
assumptions about what bound state it forms, the definition of its mass,
and the gluino fragmentation provided that it does not decay inside the
detectors.  
We believe that the
currently available data set is much more sensitive to a possible
additional contribution from the light gluino.\footnote{A paper by 
ALEPH \cite{ALEPHnew} came out
after the completion of this work, which claims to exclude a
light gluino below 6.3~GeV by combining the four-jet angular variables
with the two-jet rate.  This type of analysis may be more sensitive
to the NLO corrections.}  We argued that the
NLO corrections are unlikely to modify the conclusion;
still, this assertion needs to be justified by explicit calculations in
the future.
As a by-product of this
analysis, we discussed the effect of finite bottom quark mass on BZ and
NR distributions in detail, which is not negligible when extracting QCD
color factors at current precisions.  

\section*{Acknowledgments}
HM thanks Mike Barnett, John Ellis, Lance Dixon, Ian Hinchliffe, and
Kam-Biu Luk for useful 
conversations.  We thank Axel Kwiatkowski for comments on the manuscript.  
AdG was supported by CNPq (Brazil), and HM was supported
in part by the Director, Office of Energy  
Research, Office of High Energy and Nuclear Physics, Division of High 
Energy Physics of the U.S. Department of Energy under Contract 
DE-AC03-76SF00098 and in part by the National Science Foundation under 
grant PHY-95-14797.

\end{document}